\documentclass{optica-article}

\journal{opticajournal} 

\articletype{Research Article}
\usepackage{siunitx}
\usepackage{tabularx}
\usepackage{multirow}
\usepackage{longtable}
\usepackage{lineno}

\begin{document}

\title{Fabrication and characterization of optical filters from polymeric aerogels loaded with diamond scattering particles}

\author{Alyssa Barlis,\authormark{1,*} Haiquan Guo,\authormark{2} Kyle Helson,\authormark{1,3} Charles Bennett,\authormark{4} Carol Yan Yan Chan,\authormark{4} Tobias Marriage,\authormark{4} Manuel Quijada,\authormark{1} Ariel Tokarz, \authormark{5} Stephanie Vivod, \authormark{5} Edward Wollack, \authormark{1} and Thomas Essinger-Hileman\authormark{1}}

\address{\authormark{1} NASA Goddard Space Flight Center, Greenbelt, MD 20770, USA\\
\authormark{2} Universities Space Research Association, 425 3rd Street SW
Suite 950, Washington, DC 20024, USA\\
\authormark{3} University of Maryland, Baltimore County, 1000 Hilltop Circle, Baltimore, MD 21250, USA\\
\authormark{4} Johns Hopkins University, Department of Physics \& Astronomy, 3400 N. Charles Street, Baltimore, MD 21218 USA\\
\authormark{5} NASA Glenn Research Center, 21000 Brookpark Rd, Cleveland, OH 44135 USA}

\email{\authormark{*}alyssa.barlis@nasa.gov} 


\begin{abstract*} 
We have developed a suite of infrared-blocking filters made by embedding diamond scattering particles in a polyimide aerogel substrate. We demonstrate the ability to tune the spectral performance of the filters based on both the composition of the base aerogel material and the properties of the scattering particles. We summarize the fabrication, optical modeling, and characterization of these filters. We investigate two polyimide base aerogel formulations and the effects of loading them with diamond scattering particles of varying sizes and relative densities. We describe a model for the filters' behavior using a combination of Maxwell Garnett and Mie scattering techniques. We present optical characterization results for diamond-loaded aerogel filters with cutoff frequencies (50\% transmittance) ranging between 2.5 and 15\,THz, and confirm that the measured spectral performance is in agreement with our optical models. We also measure the filters' refractive indices in the microwave and report findings in agreement with Maxwell Garnett model predictions (typically $n < 1.08$). 
\end{abstract*}

\section{Introduction}

Aerogels are low density, highly porous, open-celled materials derived from a gel in which the liquid is removed while maintaining the self-assembled three-dimensional structure, generally by supercritical drying\cite{Aegerter2023}. The development of  silica aerogels in the 1930’s by Kistler\cite{Dorcheh2008} was later followed by the development of organic aerogels, such as cellulose, resorcinol-formaldehyde \cite{Pekala1989}, poly(vinyl alcohol) (PVA)-based \cite{Chen2014}, poly(vinyl chloride)-based \cite{Yamashita2003}, and polyimide aerogel \cite{Wendall2004, Chidambareswarapattar2010, Guo2011, Kawagaishi2007}. Among the wide array of organic aerogels, polyimide aerogels have characteristics and properties that are well-suited for use in extreme environments. Polyimides are used as high performance polymers in many aerospace applications such as structural components of aircraft, ablators for thermal protective systems, adhesives, and electrical insulation \cite{Hergenrother2003, Meador1998}. Polyimide aerogels combine the benefits of  polyimide polymers (high thermal stability, mechanical strength, high glass transition, and chemical resistance), with the benefits of aerogels (low density, nanoscale pore size and high internal surface area). 

In order to target specific applications, the properties of polyimide aerogels, including Young’s modulus \cite{Meador2012-2}, hydrophobicity\cite{Guo2012, Li2018}, optical transparency\cite{Vivod2020}, dielectric constant\cite{Meador2014}, and flexibility\cite{Pantoja2019, Guo2020, Cashman2020}, can be tailored by tuning the underlying molecular backbone structure. In addition, polymer aerogels can be formed or molded into a variety of shapes and sizes depending on the desired application. As a result, polyimide aerogels have already found use in a wide variety of applications including fire retardancy  \cite{Li2021}, thermal insulation \cite{Feng2016}, antenna substrate\cite{Meador2012}, and electrical insulation\cite{Guo2019}. 

Due to the high porosity of aerogels, they can serve as a host system to other materials without significantly compromising their performance as lightweight, low density, low permittivity materials. The available free volume of the aerogel material allows for the incorporation of particles of higher-index material during fabrication (see Section \ref{sec:synthesis}). The resulting composite material consists of a network of high-index particles held in suspension by the low-index aerogel\cite{TEH2020}. Such particle-loaded polymeric aerogel composites can subsequently be used as optical low-pass filters in the far-infrared through microwave regimes. Incoming light in the passband is transmitted through the low-reflectance and low-loss base medium while unwanted higher-frequency light is scattered and reflected by the composite aerogel. There are a variety of materials suitable to load the aerogel medium to produce short-wavelength scattering (\textit{e.g.} insulating ceramics, high-resistivity silicon, amorphous glass). Diamond was chosen because it has low electromagnetic absorption, high dielectric contrast with respect to vacuum in the microwave through the visible spectral ranges and is commercially available in a wide range of particle sizes \cite{Thomas1994, Barlis2022}.

Herein, we demonstrate the fabrication, characterization, and optical performance of diamond-loaded aerogel scattering filters. Polyimide aerogels with mechanical flexibility and robustness were fabricated as matrices to embed various sizes and concentrations of diamond particles to tune the cutoff frequency and out-of-band blocking performance of the composite material. Our filters are targeted for use in a variety of applications, from ground-based cryogenic telescope experiments to space-based planetary science probes. Section \ref{sec:fabrication} describes the fabrication process used to produce the base aerogel and composite diamond-loaded scattering filters, along with the structural and mechanical characterization of the filter material. Section \ref{sec:modeling} details the approach used to model the effect of the diamond content on the optical performance of the filters. Section \ref{sec: Characterization} reports the optical characterization of several representative filter samples and compares the modeled optical performance to the measured results.

\section{Fabrication} \label{sec:fabrication}

\subsection{Synthesis of polyimide aerogel film embedded with diamond particles} \label{sec:synthesis}

The polyimide aerogels were prepared using a combination of monomeric precursors which include biphenyl-3,3’,4,4’-tetracarboxylic dianhydride (BPDA) and a combination of diamines: 2, 2’-dimethylbenzidine (DMBZ) and 4,4’-Bis (4-aminophenoxy) biphenyl (BAPB) or DMBZ and 1,12-dodecyldiamine (DADD). The oligomers were then crosslinked using 3,5-triaminophenoxybenzene (TAB) in N-methyl-2-pyrrolidinone (NMP) followed by chemical imidization. Traditional polyimides utilize thermal imidization methods \cite{Meador1998} for ring closure of the amic acid intermediate at elevated temperatures. In this study, chemical imidization \cite{Guo2011, Meador1998, Meador2012, Guo2019, Meador2012-2, Guo2012,Li2018, Vivod2020, Meador2014, Pantoja2019, Guo2020, Cashman2020} was chosen due to its cost effectiveness and safer handling environment whereby imidization occurs using a water scavenger (acetic anhydride) and a base catalyst (triethylamine) processed at room temperature. 

\begin{figure}[ht!]
\centering\includegraphics[width=\textwidth]{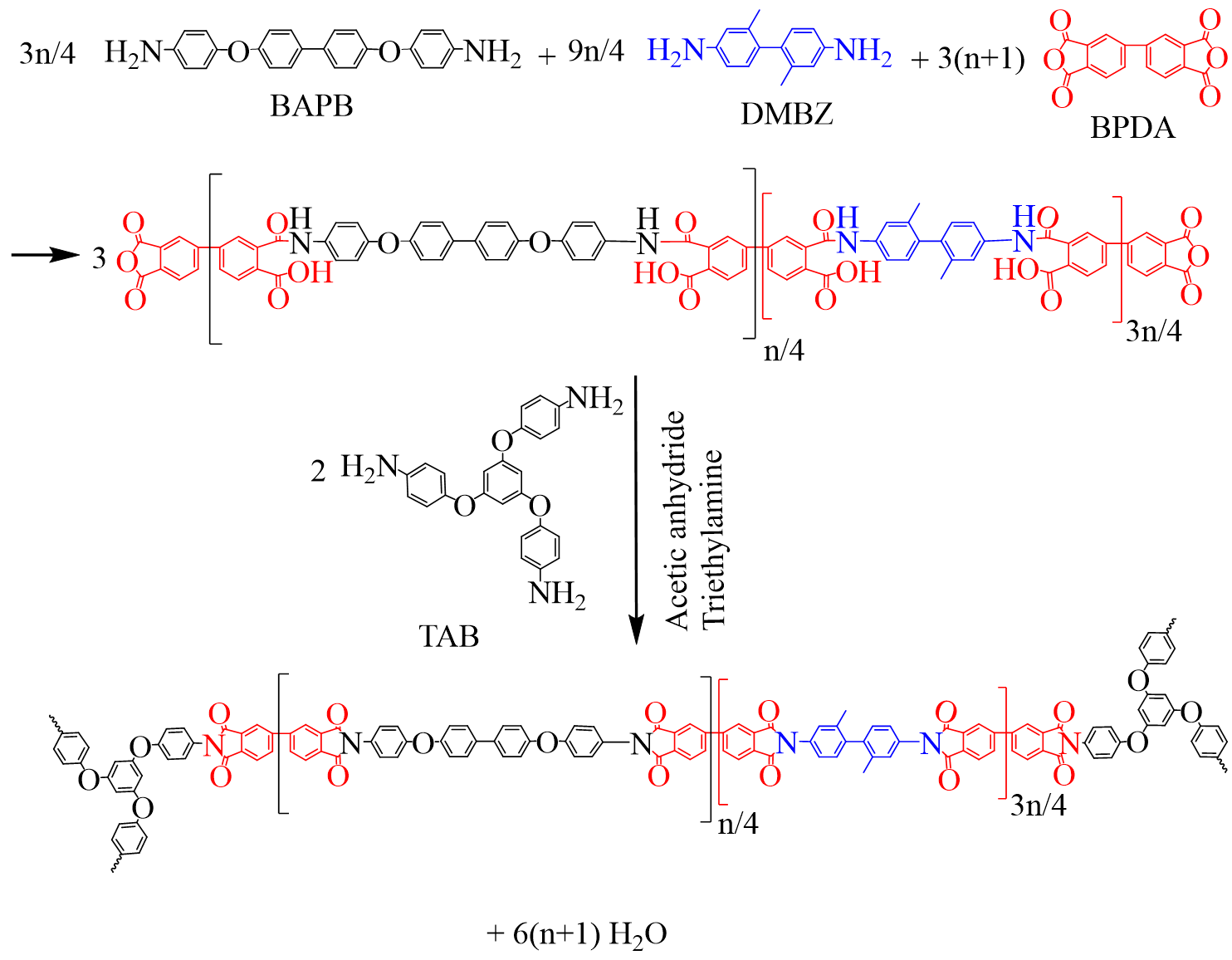}
\caption{The reaction of the polyimide aerogel network.}
\label{fig:chemical-reaction-scheme}
\end{figure}

Polyimide synthesis using BPDA as the dianhydride, a combination of diamines, DMBZ and BAPB, and a cross-linking monomer, TAB, is depicted in Fig. \ref{fig:chemical-reaction-scheme}. The diamond particle size and loading content used for each formulation are listed in the Supplement 1. For Runs 1--35, the polymer concentration was formulated at 10 wt\% and molar ratio of dianhydride to the total diamine fraction was 26:25, with a molar ratio of DMBZ to BAPB at 3:1. For example, in Run 1, 6.94\,g BAPB and 23.07\,g BPDA were mixed in 320.20\,mL NMP and stirred with a magnetic stirrer for 30 min. Then a solution of 12.01\,g DMBZ in 20.00\,mL NMP solution was added while continuing stirring until the sol was clear and an intermediate amic acid is formed, which is followed by addition of the crosslinker (0.80\,g TAB in 10.00\,mL NMP). At this stage of synthesis, the diamond particles of different sizes and concentrations were added prior to chemical imidization using 59.30\,mL acetic anhydride and 10.93\,mL triethylamine. After stirring for 15\,min, the sol was poured into molds and aged over 24 hours to ensure full gelation. Molds include 52.5\,cm $\times$ 52.5\,cm vertical molds with adjusted spacer thickness, a 3-inch diameter round silicone mold, and 20\,mL syringe molds with the tips removed. Films were also created by casting the sol on a polyethylene terephthalate (PET) carrier film on a benchtop film casting system (ChemInstruments EZ coater EC100) using a doctor blade with the gap set to the desired thickness. The resulting gels were transferred to a solvent bath of 50\% NMP + 50\% acetone, followed by 100\% acetone solvent swap four times per day. The polyimide gels undergo supercritical fluid extraction using liquid CO\textsubscript{2} to obtain the final aerogels. 

As part of this study, DADD-based aerogels were also fabricated using the same method but with 9wt\% polymer concentration, a repeat unit of n = 60, and a molar ratio of DADD:DMBZ = 3:2. DADD-based polyimide aerogel shows similar matrix properties as BAPB-based aerogel; however, this work mainly focused on BAPB aerogel due to its relatively stronger mechanical properties.

\subsection{Chemical and Mechanical Characterization}
The skeletal density of the samples was measured using a Micromeritics AccuPyc 1340 helium pycnometer to determine the percent porosity, calculated as:

\begin{equation}\label{porosity}
    \rm{Porosity} = 100 \% \times (1- \frac{\rho_{\rm{b}}}{\rho_{\rm{s}}}),
\end{equation}

\noindent where $\rho_{\textrm{b}}$ is the bulk density and $\rho_{\textrm{s}}$ is the skeletal density measured by helium pycnometry \cite{Ayral1992}. Polyimide aerogel samples were sputter coated with a 5\,nm layer thickness of platinum to obtain scanning electron micrographs, using a Hitachi S-4700 field emission scanning microscope. Brunauer-Emmett-Teller (BET) surface area was measured with a Tristar II 3020 Surface Area/Pore Distribution analyzer (Micromeritics Instrument Corp.). Before running nitrogen-adsorption porosimetry, the samples were out-gassed at 80\,$^\circ$C overnight. Thermal conductivities of the samples were measured on a calibrated hot plate device (designed by Aerogel Technologies, LLC) using a flat monolithic sample in direct contact with a plate of National Institute of Standards and Technology (NIST) reference material of similar planar dimensions with known thermal conductivity. The sample and reference material were sandwiched between an aluminum plate maintained at 37.5\,$^\circ$C and a flat-bottomed aluminum vessel filled with an ice bath. Temperatures were measured at the hot side, cold side, and at the interface between the two materials. The heat flux through the material was calculated from the temperature drop across the NIST standard. This heat flux was then used with the thickness and temperature drop across the unknown sample to calculate the sample’s thermal conductivity. Optical microscope images of polyimide aerogel films were captured using a Nikon SMZ 1270 optical microscope.

\subsection{Results and Discussion}

\begin{figure}[h]
    \centering
    \includegraphics[width = \textwidth]{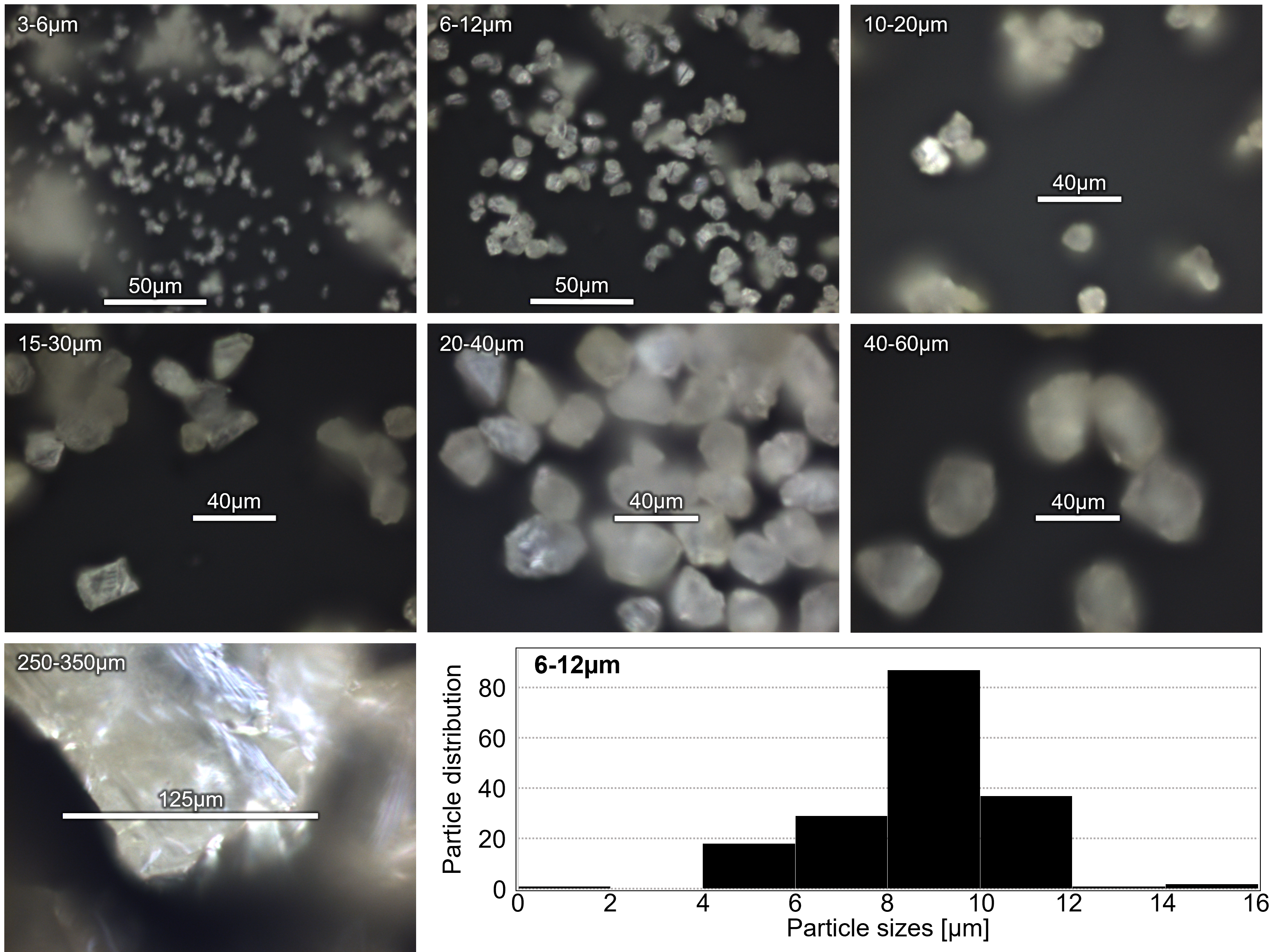}
    
    \caption{Optical microscope images of some of the diamond particle samples. Per the manufacturer, the distribution of particle sizes within each batch is approximately Gaussian. The size ranges are specified according to 3-$\sigma$ bounds of these Gaussian distributions. }
    \label{fig:diamond_particles}
\end{figure}

Fig. \ref{fig:diamond_particles} shows optical microscope images of the diamond particles prior to being incorporated within the polyimide network. The diamond particles have irregular shapes and come in a range of sizes. The diamond particle sizes used for this work include: 3--6\,\textmu m (median 4.2\,\textmu m), 6--12\,\textmu m (median 9\,\textmu m), 10--20\,\textmu m (median 15\,\textmu m), 15--30\,\textmu m (median 22.5\textmu m), 20--40\,\textmu m (median 30\,\textmu m), 40--60\,\textmu m (median 47\textmu m), and 250--350\,\textmu m (median 300\,\textmu m). Particle size was verified using optical microscopy by averaging over the measured lengths of the primary axes. 

\begin{figure}[]
\centering\includegraphics[width=0.66\textwidth]{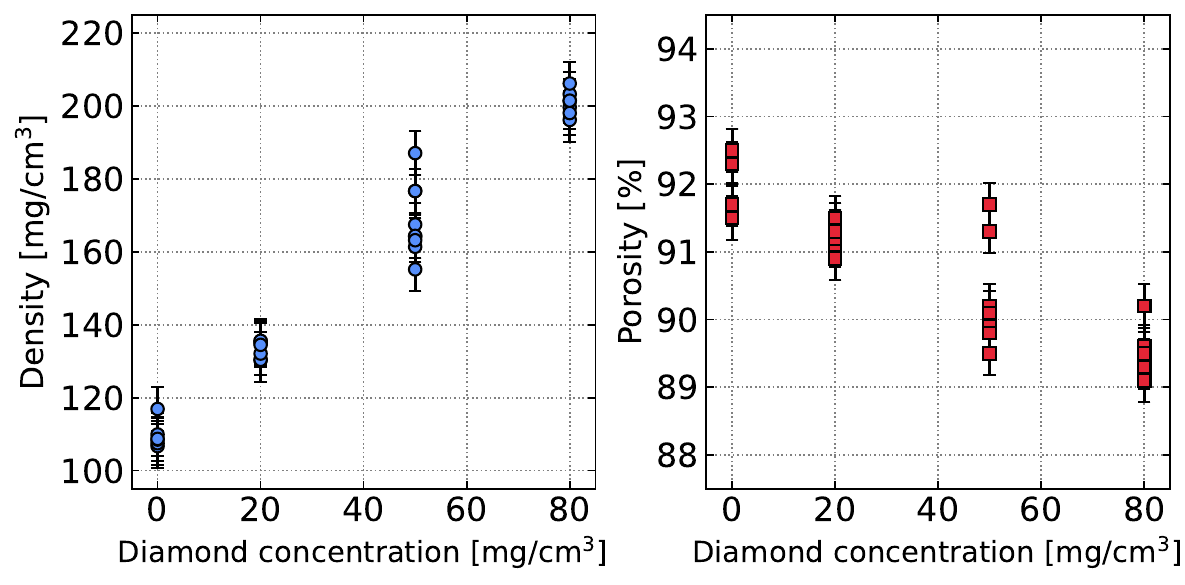}
\caption{Plots of measured density (\textit{left}) and porosity (\textit{right}) as a function of diamond concentration of samples of polyimide aerogel composite with diamond particles. Measured density scales linearly with diamond concentration, confirming that the aerogel matrix remains largely unchanged as particles are added. Measured composite porosity decreases slightly as a function of diamond concentration due to the volume taken up by the diamond particles. }
\label{fig:density-and-porosity-vs-DiamondConcentration}
\end{figure}

The properties of the polyimide aerogel/diamond composites are shown in the Supplement 1. Each run describes a formulation of polyimide aerogels or polyimide aerogel composite with diamond particles. The left panel of Fig. \ref{fig:density-and-porosity-vs-DiamondConcentration} shows the measured density of the polyimide aerogel versus diamond concentration. 
The density of the polyimide aerogel increases approximately linearly with diamond particle concentration, going from 100\,mg/cm$^3$ to above 200\,mg/cm$^3$. The linear shrinkage of the polyimide aerogel composites ranges from 6.4--14.1\% and appears uniform in all dimensions and independent of diamond particle size or concentration. Overall, all the polyimide aerogels show thermal conductivity ranging from 23.9--41.3\,mW/(m$\cdot$K).

\begin{figure}[]
\centering\includegraphics[width=\textwidth]{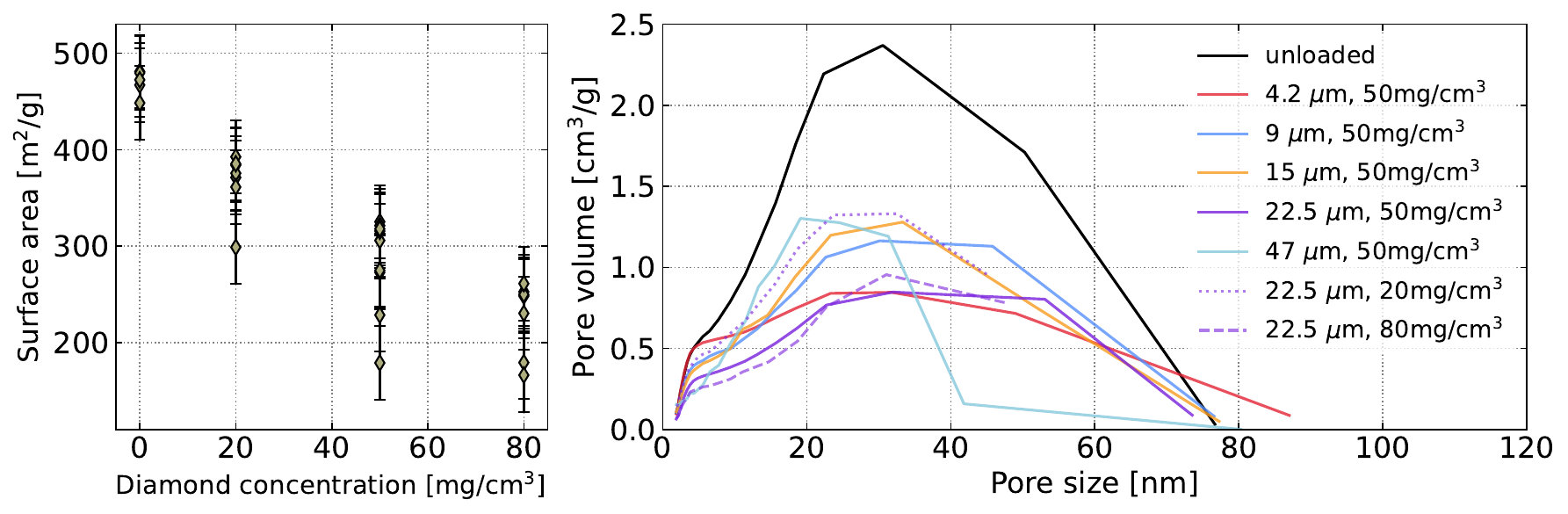}
\caption{\textit{Left}: Measured surface area of polyimide aerogel composites as a function of total diamond concentration. \textit{Right}: Pore size distribution (pore volume per mass) of polyimide aerogel (black line) in comparison to polyimide aerogel diamond composites with a range of diamond loading compositions (colored lines). The decrease in surface area and pore volume as particles are added is largely due to the increase in composite density with diamond particle concentration and not indicative of a change in aerogel pore structure.}
\label{fig:SurfaceArea_PoreVolume}
\end{figure}

The right panel of Fig. \ref{fig:density-and-porosity-vs-DiamondConcentration} shows the measured porosity of the polyimide aerogel versus diamond concentration. As the diamond concentration increases to 80\,mg/cm$^3$, the porosity of the aerogel composites slightly decreases from 92\% to 89\%. This decrease is largely explained by the volume taken up by the diamond particles in the composite, which is approximately 2.3\% at 80\,mg/cm$^3$, indicating that the aerogel pore structure is not strongly affected by the presence of the diamond particles. The mesoporous/macroporous polyimide network of the aerogels, with typical pore sizes of 5--100~nm, wraps around the much-larger ($> 1$~\textmu m) diamond particles. 

The left panel of Fig.~\ref{fig:SurfaceArea_PoreVolume} shows the measured surface area per unit mass versus diamond concentration. The impact of concentration is notably pronounced, and the surface area drops from $\sim 480$\,m\textsuperscript{2}/g to $\sim 210$\,m\textsuperscript{2}/g when the diamond concentration increases to 80\,mg/cm$^3$. This is due to the increase in total density from $\sim 110$\,mg/cm$^3$ to $\sim 200$\,mg/cm$^3$ as diamond concentration increases to 80\,mg/cm$^3$ due to the inclusion of the diamond particles (see Fig.~\ref{fig:density-and-porosity-vs-DiamondConcentration}) rather than changes in the pore structure itself. The right panel of Fig. \ref{fig:SurfaceArea_PoreVolume} shows the measured pore size distributions, in units of pore volume per mass, within the polyimide aerogel composites, obtained from N\textsubscript{2} adsorption porosimetry. The overall decrease in pore volume as particles are added is again consistent with the increase in composite density due to inclusion of the diamond particles. The distribution of pore sizes is largely unchanged when diamond particles are added, with an unexplained exception for the sample with 50\,mg/cm$^3$ of relatively large (47~\textmu m) particles. 

\begin{figure}[h]
    \centering
    \includegraphics[width = \textwidth]{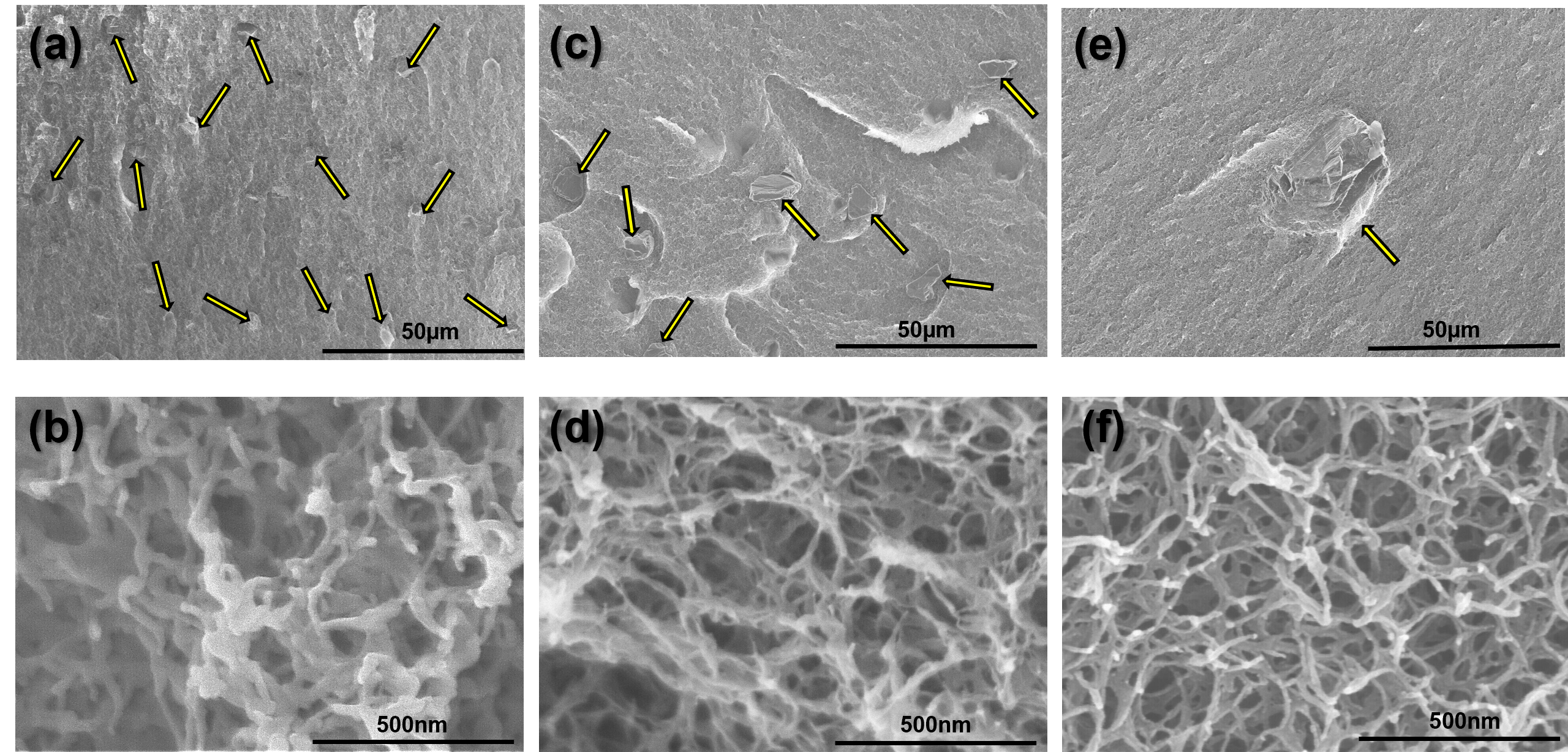}
    
    \caption{Scanning electron micrograph images of three samples of diamond-loaded aerogel scattering filters. Yellow arrows indicate the positions of diamond particles in the top row of images. Each of the samples shown is a polyimide aerogel loaded with one size range of diamond scattering particles at a concentration of 50\,mg/cm$^3$: (a) and (b) 4.2\,\textmu m; (c) and (d) 9\,\textmu m; and (e) and (f) 47\,\textmu m.}
    \label{fig:SEM}
\end{figure}

Fig.~\ref{fig:SEM} shows scanning electron microscope (SEM) images of three polyimide aerogel composites loaded with diamond particles of different sizes at 50\,mg/cm$^3$ loading density. Diamond particles are visible on the surface of the polyimide aerogel network at low magnification (Fig. \ref{fig:SEM} a, c, and e). Higher magnification SEM images (Fig. \ref{fig:SEM} b, d, and f) in regions away from diamond particles show the fiber connected networks within the polyimide aerogels. The SEM images also verify that the mesoporosity does not change with particle incorporation. 

\begin{figure}[h]
    \centering
    \includegraphics[width = \textwidth]{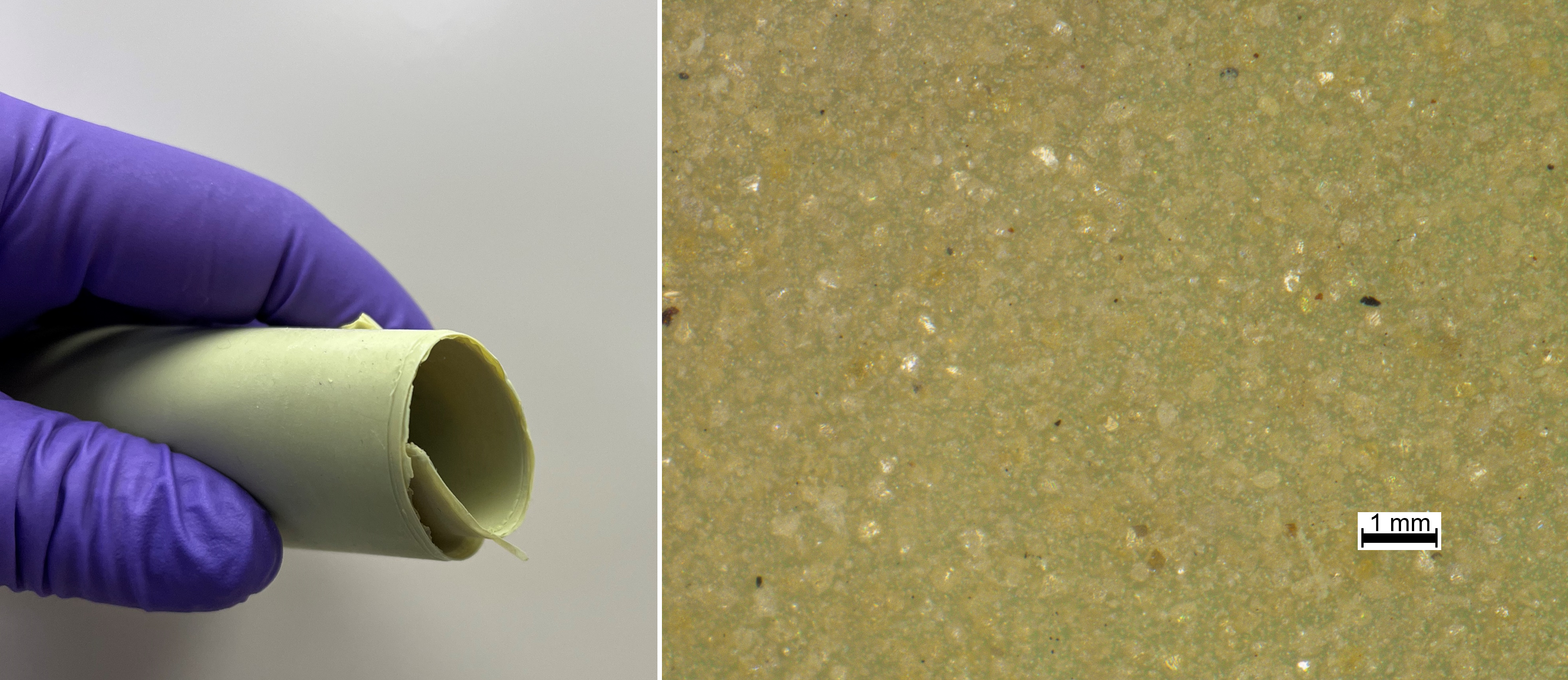}
    
    \caption{\textit{Left}: Polyimide aerogel scattering filter prepared as a 1mm thick film with diamond scattering particles of three size ranges, 3--6\,\textmu m , 10--20\,\textmu m , and 40--60\,\textmu m (run 32 as described in Supplement 1). \textit{Right}: Optical microscope image of the same filter.}
    \label{fig:OpticalImages}
\end{figure}

Formulations containing multiple sizes and concentrations of diamond particles are shown in Supplement 1. Combinations of diamonds of 3--6\,\textmu m, 10--20\,\textmu m, 15--30\,\textmu m, or 40--60\,\textmu m were incorporated into the aerogels in Runs 30--35 with homogeneous diamond distribution. As seen in the left panel of Fig. \ref{fig:OpticalImages}, a 1\,mm thick polyimide aerogel film prepared with three diamond particle sizes, 3--6\,\textmu m, 10--20\,\textmu m, and 40--60\,\textmu m, is flexible enough to be rolled up. The diamond particles are homogeneously distributed throughout the filter, as shown in the optical microscope image in the right panel of Fig. \ref{fig:OpticalImages}. In run 36, the larger diamond particles with 250--350\,\textmu m settled at the bottom of the aerogels due to gravity. The bottom regions of these samples show lower surface area 177.8\,m\textsuperscript{2}/g than the top, 309.2\,m\textsuperscript{2}/g.

\section{Modeling} \label{sec:modeling}
Aerogels are highly porous materials with typical pore sizes less than 100\,nm. The electromagnetic behavior of the aerogel can be thought of as a homogeneous effective medium and modeled using Maxwell Garnett theory\cite{MaxwellGarnett, Niklasson1981, Sihvola2008, Markel2016} at wavelengths longward of the infrared. The base aerogel is modeled as a mixture of polymer and air/vacuum, and a loaded aerogel is modeled as a mixture of the base aerogel with embedded diamond particles. Because the base polyimide polymer has a moderate index of refraction ($n = 1.77 + 0.004i$ at 167\,\textmu m \cite{Smith1975}) and low volume-filling fraction, the resulting aerogel has a very low index of refraction ($n \leq 1.1$). For the diamond-loaded aerogel filters described in Section \ref{sec:fabrication}, the contrast between the low index of refraction base aerogel and the higher index of refraction from the embedded diamond particles in the composite medium produces Mie scattering. Mie scattering of light leads to rejection of light with a wavelength roughly equal or less than the size of the higher index of refraction particles \cite{TEH2020, 2022SPIEKH}.

The diamond-loaded polyimide aerogel filter behavior is well described by a Mie scattering model. PUREON Microdiamant MONO-ECO monocrystalline diamond powders\cite{Pureon} with size ranges of 1--3\,\textmu m, 3--6\,\textmu m, 6--12\,\textmu m, 10--20\,\textmu m, 15--30\,\textmu m, 20--40\,\textmu m, 40--60\,\textmu m, and 250--350\,\textmu m were chosen to cover a broad range of particle sizes while also having well-controlled distributions of particles within each range. These sizes of scattering media cover a broad range of potential cutoff frequencies in the far-infrared through millimeter wave regimes. Particle size distributions are approximately Gaussian, with the end ranges representing the approximate three standard deviation limits for particle diameter. Table \ref{powders} shows a sub-sample of particle size distribution data. Diamond powder composition, particle sizes, and shapes were later analyzed with optical microscopy and Raman spectroscopy. Fig. \ref{fig:diamond_particles} shows images of the diamond particles. 

\begin{table}[h]
\centering
\begin{tabular}{ |c|c| } 
 \hline
 \textbf{Powder Size Range} &  \textbf{Median Size} \\ 
 \hline
 3-6\,\textmu m & 4.2\,\textmu m \\
 \hline\
 6-12\,\textmu m & 9\,\textmu m \\
 \hline
 15-30\,\textmu m & 22.5\,\textmu m \\
 \hline
 40-60\,\textmu m & 47\,\textmu m \\
 \hline
\end{tabular}
\caption{Diamond particle size distributions as specified by the manufacturer and corresponding median particle sizes as measured by optical microscopy. All diamond powders are from the PUREON Microdiamant MONO-ECO monocrystalline synthetic diamond. Median particle sizes all fall within the stated specifications from the manufacturer.}
\label{powders}
\end{table}

The diamond particle size and loading density are controlled to tune the filter to the desired cutoff frequency. Particle size has the dominant effect on cutoff frequency. Larger particles lead to lower cutoff frequencies and smaller particles lead to higher cutoff frequencies. For a given particle size distribution, increasing the loading density results in a filter with a lower cutoff frequency and vice versa. Beyond the cutoff frequency, out-of-band blocking performance is determined by particle size, loading density, and filter thickness. In general, smaller particles reject light more efficiently than larger particles. Geometric scattering cross-section is proportional to the square of the particle radius (area $\propto r^2$), while mass density scales with to the cube of particle radius (volume $\propto r^3$). Therefore, for a given loading density, larger-radius particles present a smaller total scattering cross-section than smaller-radius particles. Low loading densities of large particles means the particles are on average very far apart and light can ``leak'' through the material. Low densities of very small particles will still have many particles present to scatter light. Increasing the thickness of the aerogel filters with the largest particles can mitigate these effects and improve out-of-band performance.

\begin{figure}[h]
    \centering
    \includegraphics[width = \textwidth]{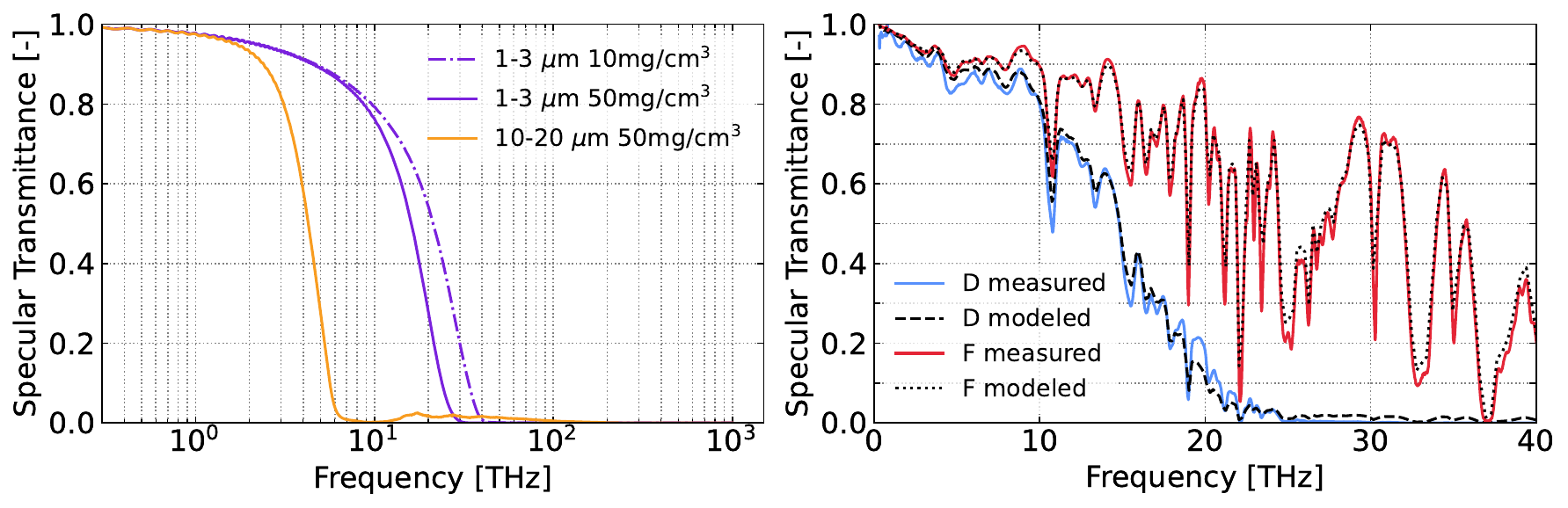}
    \caption{\textit{Left}: Demonstration of the effects of scattering particle loading density and radius on modeled aerogel filter cutoff frequency. All three curves show the behavior of a 1\,mm thick sample. Peaked features in the orange model above the cutoff frequency are artifacts of modeling and are not present in measured samples. \textit{Right}: Modeled and measured specular transmittances near the cutoff for samples D and F as described in Table \ref{table:params}.}
    \label{fig:comparison}
\end{figure}

The base polyimide aerogel formulation described in Section \ref{sec:synthesis} has an average density of 110\,mg/cm$^3$, and we have fabricated aerogel scattering filters with a range of diamond particle loading densities ranging from 20\,mg/cm$^3$ up to 100\,mg/cm$^3$. Fig. \ref{fig:comparison} demonstrates the effects of varying particle radius and density on predicted filter cutoff frequency for several filter examples. Multiple particle size distributions and densities can be mixed together to allow simultaneous tuning of cutoff frequency and out-of-band blocking performance. For example, although samples D and F contain the same total particle density, their spectral behavior depends strongly on the difference in particle sizes used (see Table \ref{table:params} for formula description). 

In order to improve upon the approach in \cite{TEH2020}, the model has been refined to incorporate multiple scatter size distributions, with different statistical profiles, including Gaussian, uniform, and skew profiles. By combining larger particles with smaller particles, we can reduce or eliminate any ``blue light leak'' above the desired cutoff frequency. For simplicity, the model presented in \cite{TEH2020} used a constant dielectric function for the base aerogel. Building upon our optical characterization discussed in Section \ref{sec: Characterization} we have incorporated the properties of the unloaded aerogel performance into the model. To capture the absorption and emission bands of the polymer chains, spectral measurements of the transmittance and reflectance are captured using the techniques described in Section \ref{sec: Characterization}. These data are used to build a dielectric function for the base aerogel matrix across a wide range of infrared frequiencies. As Fig. \ref{fig:Tspec_demo_cutoff_loaded-vs-unloaded} shows, many molecular absorption features appear in the transmittance spectra of unloaded BAPB-based and DADD-based aerogel films between 10 and 100\,THz. T With this knowledge we can more accurately predict the filter performance and tune filter performance to specific mission requirements as desired. We can also subtract the effects of the base aerogel model to confirm that the behavior of the added scattering particles is Mie-like in reality. With this iterative approach, we are able to closely model the optical behavior of our aerogel scattering filters.

\section{Optical Characterization}
\label{sec: Characterization}
\subsection{Facilities} 
\label{sec:facilities}
To fully characterize the optical performance of the aerogel scattering filters, we measure the samples' transmittance and reflectance over as broad a spectral range as possible. We use several instruments to do this. 

The PerkinElmer LAMBDA 950 instrument is a dual-beam monochromator that operates at optical wavelengths of 200--3300\,nm. We use a photomultiplier tube in the 200--900\,nm range and a PbS photocell detector in the 800--3300\,nm range. The instrument's spectral accuracy is $<\pm$2\,nm. The PerkinElmer instrument operates at ambient atmospheric pressure. 

The Bruker 125HR instrument is a versatile Fourier Transform Interferometer that operates at wavelengths between 500~nm and 1~mm. To span this range, we combine data taken in several instrument configurations. For wavelengths between 14~\textmu m and 1~mm, we use a mercury arc lamp as a white light source, Mylar film beam splitter, and Si bolometer cooled to 4.2~K as a detector. For wavelengths 2--14~\textmu m, we use a Globar white light source, KBr beam splitter, and room temperature deuterated L-alanine doped triglycine sulphate (DTGS) detector. In addition to a basic sample mount translation stage for measuring specular transmittance, we utilize an accessory stage for measuring specular reflectance at 10$^{\circ}$ angle of incidence referenced to a gold mirror. The spectral resolution of the Bruker 125HR is 0.0063~cm$^{-1}$, while the photometric precision is $\pm$1\%. All measurements with the Bruker are performed under vacuum (pressure below 0.4\,T).

Finally, we use an Agilent PNA-X vector network analyzer (VNA) to perform free-space two-port reflectance measurements. The measurement setup and analysis procedure are discussed in \cite{Eimer2011}. The VNA measurements allow us to characterize the samples in waveguide between 25 and 50\,GHz and quasi-optically between 75 and 300~GHz. Specifically, reflectance measurements at 140--220\,GHz were chosen because this frequency band contains more than three fringe features for samples of order $\sim$10\,mm thick for the range of expected index of refraction, $n\simeq1.05-1.1$, for diamond-loaded polyimide aerogels. 

\subsection{Aerogel Optical Performance}

\begin{figure}[]
\centering\includegraphics[width=\textwidth]{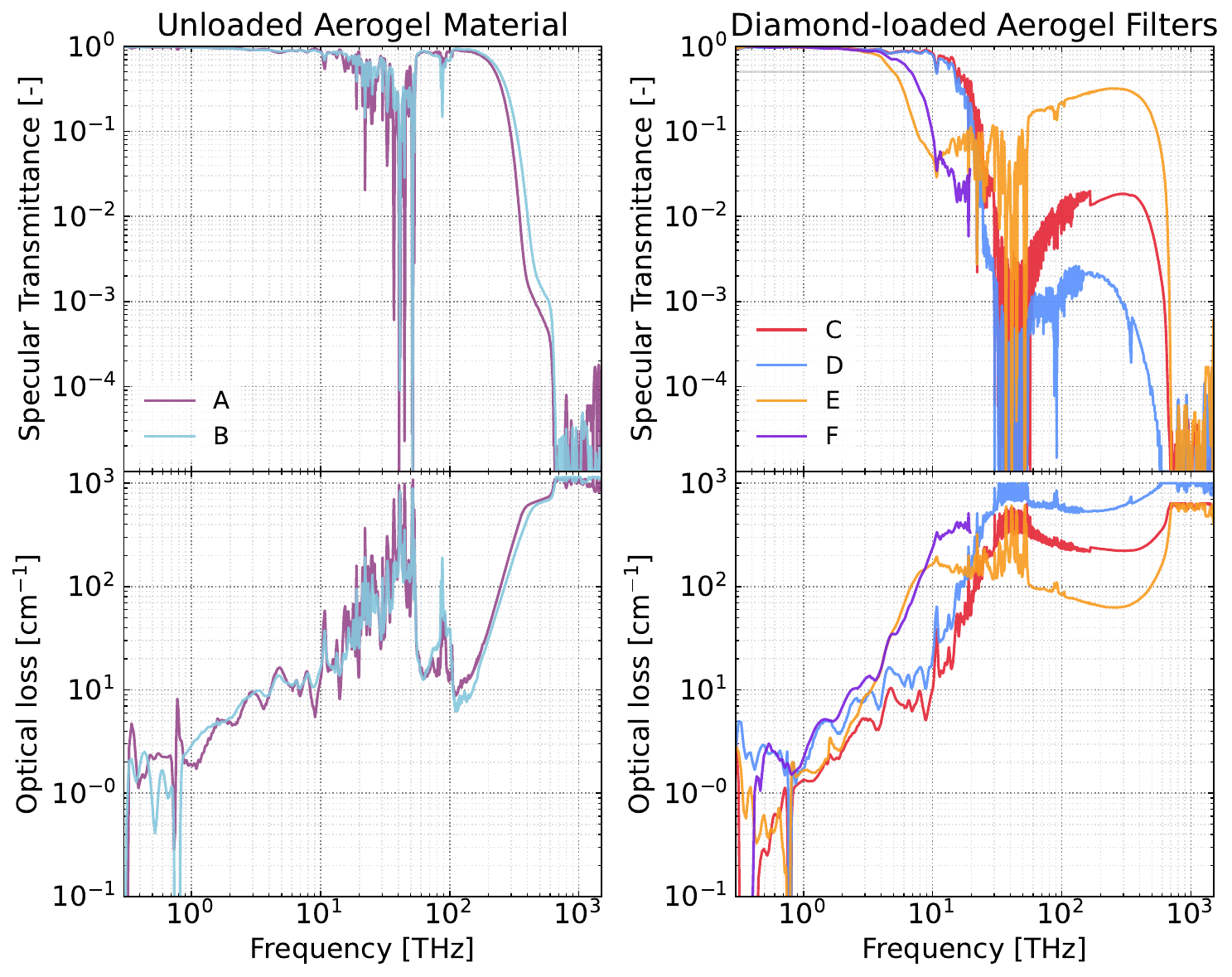}
\caption{Far-infrared specular transmittance spectra and optical loss of unloaded polyimide aerogel material as well as representative samples of aerogel scattering filters. For descriptions of each sample, see Table \ref{table:params}. Thicknesses of measured samples are as follows: A: 0.11~mm, C: 0.18~mm, D: 0.12~mm, E: 0.18~mm, F: 0.1~mm. (Color scheme from \cite{Petroff2021}.)}
\label{fig:Tspec_demo_cutoff_loaded-vs-unloaded}
\end{figure}

\begin{figure}[]
\centering\includegraphics[width=\textwidth]{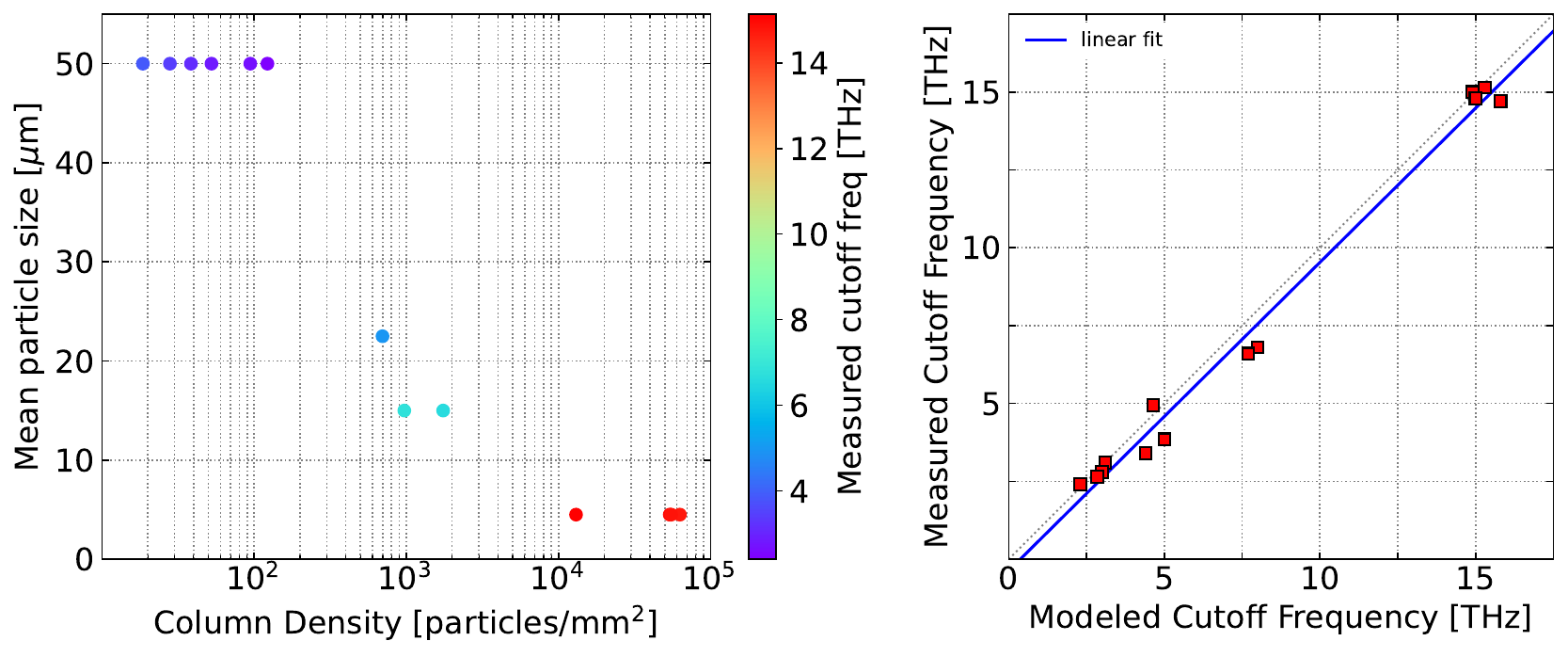}
\caption{Cutoff frequency parameter space and measured versus modeled cutoff frequency (50\% transmittance point) for diamond-loaded polyimide polymer aerogel filter samples. \textit{Left}: Explored parameter space of diamond particle size, loading density (parameterized as column density), and measured cutoff frequency. Particle size has the strongest effect on cutoff frequency but varying the loading density can allow for fine-tuning of the desired cutoff frequency. \textit{Right}: There is good agreement between the measured cutoff frequencies for a variety of aerogel filter samples and the Mie-scattering modeled cutoff frequencies. Accurate modeling allows for the design of filters to meet specific filter parameters.}
\label{fig:cutoff_frequency_combined-figure}
\end{figure}

As a result of our improvements to the Mie scattering model and the tightly-controlled particle size distributions, we are able to design and fabricate filters that span a wide range of parameter space. Table \ref{table:params} shows a selection of different mean particle sizes, loading densities, and subsequent cutoff frequencies available to us. We define the cutoff frequency for a filter as the 50\% transmittance point. Larger particles can set cutoff frequencies as low as 2.5\,THz, while smaller particles produce cutoff frequencies at approximately 20\,THz.  Fig. \ref{fig:Tspec_demo_cutoff_loaded-vs-unloaded} shows transmittance and optical loss measurements of six different aerogel and composite samples. Optical loss is calculated as: 

\begin{equation}\label{loss}
    {\alpha}_\nu = \frac{-\ln{({T}_\nu)}}{d},
\end{equation}

\noindent where $\alpha_\nu$ is the loss per unit length at frequency $\nu$, $T_\nu$ is the specular transmittance at frequency $\nu$ and $d$ is the thickness of the sample. To ensure calculated loss values remain physically-feasible, the lower limit of $T_\nu$ is set at $1\times 10^{-5}$ based on the noise floor of our highest-frequency detectors. Although two different polyimide aerogel base formulations were investigated, we chose the BAPB-based formulation described in Section \ref{sec:fabrication} for the remaining scattering filters we characterized optically because the unloaded BAPB sample shows slightly stronger mechanical properties and higher transmittance below 100\,THz. The right panel of Fig. \ref{fig:cutoff_frequency_combined-figure} shows the measured versus modeled cutoff frequencies for a range of samples. Broadly, we find good agreement between modeled and measured transmittance spectra (including cutoff frequencies) for all of the samples we studied. 

\begin{figure}[h]
    \centering
    \includegraphics[width = 0.75\textwidth]{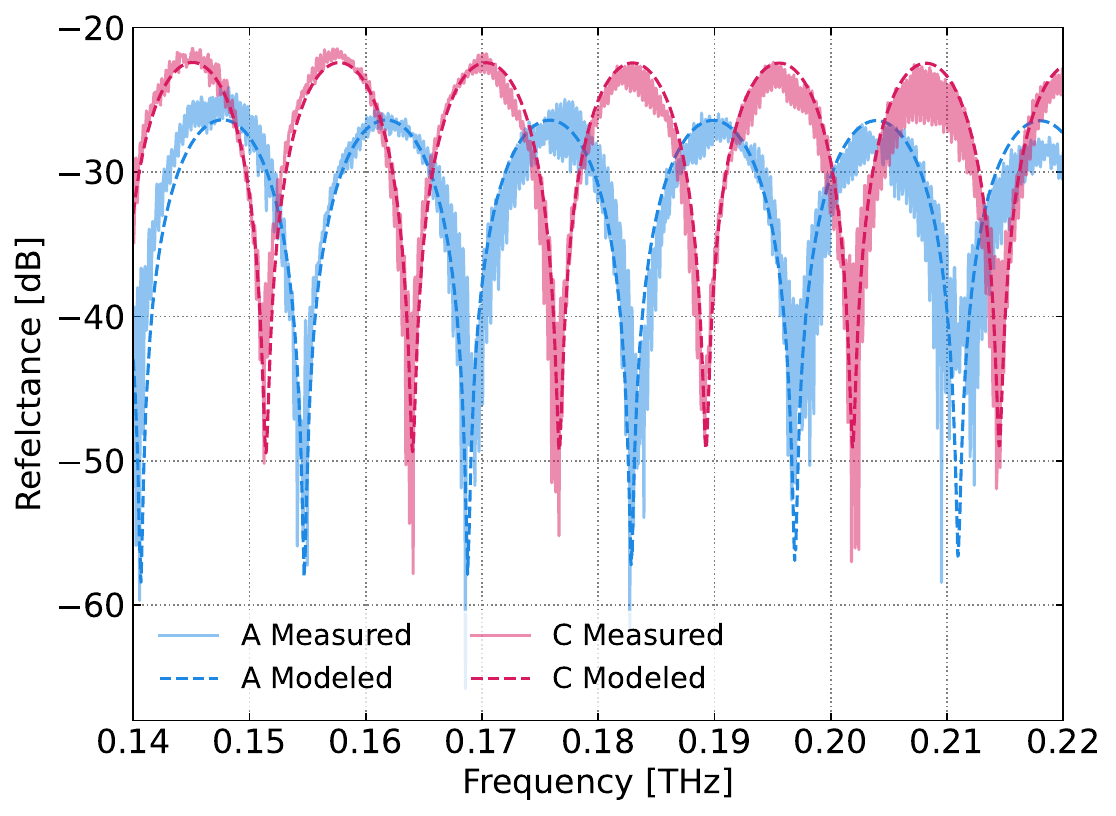}
    \caption{Vector Network Analyzer measurements of two different aerogel samples. Measurements were performed in free space at WR-05. The solid lines are VNA measurement data and the dashed lines are best-fit model curves used to determine the index of refraction  for each sample. Sample A is an unloaded BAPB-based polyimide aerogel. Sample C is a diamond-loaded polyimide aerogel, consisting of 3--6\,\textmu m particles at a loading density of 50\,mg/cm$^3$. Diamond-loading increases the reflectance of the sample compared to the plain aerogel, however both samples have very low reflectance from 140 to 220\,GHz. }
    \label{fig:VNA_measurements}
\end{figure}

Transmittance and reflectance of aerogel samples approximately 1\,cm thick were measured in the quasi-optical VNA testbed. Using the resultant reflectance data, we are able to constrain the index of refraction and sample thickness to 2\% or better in the microwave regime. VNA data are fit to a basic dielectric slab model. Fig. \ref{fig:VNA_measurements} shows two sets of VNA data along with fitted model curves. 

\begin{figure}[]
\centering\includegraphics[width=0.75\textwidth]{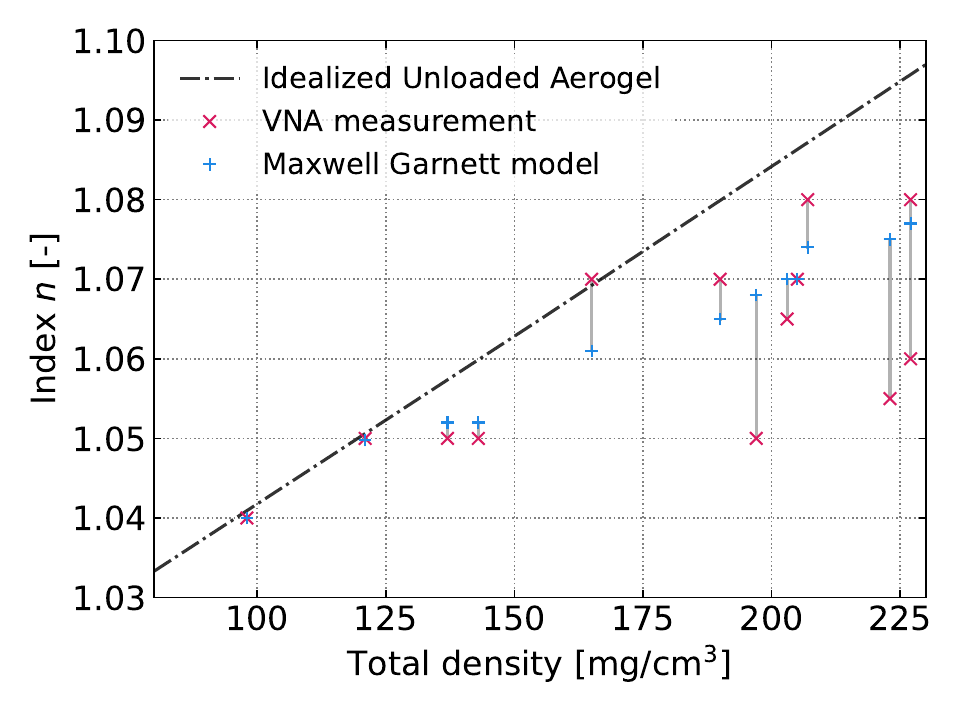}
\caption{Measured polyimide aerogel filter density versus index of refraction. Index of refraction for one sample can be determined in two different ways. Index of refraction is for wavelengths > 500\,\textmu m for the millimeter and microwave regimes. The pair of data points for the same sample are connected via a gray line. Data in blue are the calculated indices using the Maxwell Garnett effective medium approximation adopting the measured sample densities. Data in red are the sample indices determined by fitting data from the VNA to an electromagnetic model of a homogeneous dielectric slab. Black line shows the expected index for an unloaded polyimide aerogel across a range of densities. Due to sample variations in thickness and particle homogeneity, measured VNA data differ from the Maxwell Garnett method using the density measurements. Variations in shrinkage from batch-to-batch cause deviations from the idealized aerogel indices and the density-measurement derived indices.}
\label{fig:density_vs_index}
\end{figure}
A selection of sample parameters derived from VNA measurements is shown in Table \ref{table:params}. A subset of existing samples was chosen to effectively cover the range of parameter space available. We find indices of refraction of $n \leq 1.08$. Even though VNA measurements were performed from 140--220\,GHz ($\simeq$ 1.4--2.1\,mm or $\simeq$ 4.7--7.3\,icm) the index of refraction is valid across the millimeter and microwave regimes. Because of the expectedly low volume filing fraction of the constituent materials (diamond and polyimide) relative to vacuum, the samples would need to be $\sim$30 times longer to be able to appropriately measure intrinsic loss with this quasi-optical measurement strategy. Maxwell Garnett theory predicts predicts the ratio of the displacement over the conduction current or the loss tangent is to be $\tan \delta \leq 0.001$. Measured in waveguide from 26--40\,GHz, the dielectric function of the BAPB-based aerogel without diamond loading is $ \epsilon^* = 1.16 \pm 0.01 + 0.0005i \pm 0.0002i$ . 

The dielectric slab model assumes that samples are perfectly plane parallel, which is not the case for the actual samples as manufactured. By fitting the VNA data with a dielectric model \cite{Grossman95} that incorporates a decoherence factor, $g$, where $g = 1$ is perfectly coherent and $g = 0$ is perfectly incoherent, we find the samples to have a range of $0.95 < g < 0.999$. Thickness variations, deviations from planarity, (\textit{e.g.} wedged or curves samples) and structural defects like bubbles will decrease the coherence of the mm-wave light passing through the sample, producing an effect that is degenerate with higher intrinsic material losses. Similarly, the index of refraction from quasi-optical VNA measurements may differ from the homogeneous dielectric slab model because the above effects will alter the coherence of the reflections from each interface. Additionally, during polymerization, the samples typically lay with the optical axis parallel with the gravitational vector. Gravity pulls the diamond particles down to one side and the effect is increased as the particles increase in size. This creates a polymer that has a more densely loaded and a less densely loaded side. The less dense side will have a lower reflectivity than expected and vice versa. In certain applications a gradient index may be desirable when very high loading densities are needed. 

Fig. \ref{fig:density_vs_index} shows a comparison between the index of refraction for the aerogel samples in Table \ref{table:params} measured in two different ways.  The first method uses VNA reflectance data and the fitting parameters from the dielectric slab model, as mentioned above. The second uses measured densities of the dried samples to calculate an index from the Maxwell Garnett approximation using knowledge of the bulk material dielectric data. The disagreement between the two data sets in Fig. \ref{fig:density_vs_index} is due to a confluence of factors affecting both measurement techniques. All aerogels shrink during the final drying phase, increasing in density slightly from their pre-dried state. Various factors during fabrication can affect the degree to which samples shrink. Shrinkage can be anticipated and taken into account during fabrication, but cannot be fully controlled. For diamond-loaded samples, it is impossible to measure the final densities of the polyimide and diamond individually so only total density is known. Measurements of total loaded sample density are combined with knowledge about the shrinkage of unloaded polyimide aerogels to calculate estimated final densities for the diamond particles and the aerogel matrix. Ideally, fabrication of each diamond-loaded sample would be accompanied by an unloaded witness sample to enable more accurate estimation of final diamond and polyimide. Not all loaded aerogels we studied had witness samples, so measured shrinkages from the same formulation from different batches were used. This adds additional systematic error and increasing potential discrepancy between VNA measurement and density measurement derived index of refraction. Regardless of the aforementioned effects, our prototype filters generally display the desired characteristics of low index and low loss. 
\\

\begin{table}[h]
\centering
\resizebox{\linewidth}{!}{%
\begin{tabular}{|c|c|c|c|c|} 

 \hline
 \textbf{Sample} & \textbf{Cutoff} &\textbf{Index of} & \textbf{Sample} \\ 
 \textbf{Name} & \textbf{frequency}&  \textbf{refraction} & \textbf{Notes}\\
 {} &  $f$ &  $n$  & {}\\
 
 \hline
A & [-] & 1.06 & Unloaded BAPB Formula - no cutoff\\
 \hline
  B & [-] & 1.05 & Unloaded DADD Formula - no cutoff \\
 \hline
 C & 15\,THz& 1.08 & 3--6\,\textmu m particles at 50\,mg/cm$^3$\\
 \hline

   D & 14.8\,THz& 1.06  & 3--6\,\textmu m particles at 80\,mg/cm$^3$\\
 \hline

   E & 4.94\,THz& 1.08  & 15--30\,\textmu m particles at 80\,mg/cm$^3$\\
 \hline
   \multirow{2}{*}{F} & \multirow{2}{*}{6.8\,THz}& \multirow{2}{*}{1.065}  & 10--20\,\textmu m particles at 60\,mg/cm$^3$ and\\
 {} & {}& {} & 3--6\,\textmu m particles at 20\,mg/cm$^3$\\
 \hline


\end{tabular}
}
\caption{Summary of measurement data from a variety of diamond-loaded polyimide aerogel samples. Cutoff frequencies were measured on the Bruker Fourier Transform Spectrometer as described in Section \ref{sec:facilities}. The index data were derived from best-fit parameters to VNA reflectance data. }
\label{table:params}
\end{table}

\section{Conclusion}
In conclusion, we have developed a suite of IR-blocking scattering filters using polyimide aerogel material loaded with diamond scattering particles. Polyimide aerogels are a flexible, durable, and cryogenically-compatible substrate with a low index of refraction because of their high ($\sim 90$\%) porosity. Diamond particles with tightly controlled size distributions provide ideal scattering media because of their high refractive index, low cost, commercial availability, and well understood dielectric properties. Combining the two allows the construction of infrared blocking filters for far-infrared observations and measurements from the terahertz through microwave regimes. We characterized the chemical, physical, and thermal properties of the aerogel-diamond composites, finding that the pore structure of the polyimide aerogel is largely unchanged by the incorporation of up to 2.3\% diamond particles by volume (80\,mg/cm$^3$). We have demonstrated the ability to predict their optical performance using a Mie scattering model and synthesize filters with tunable frequency cutoffs between 2.5 and 15\,THz with this approach. Inclusion of larger and/or smaller particles can expand that range from $\sim0.5$\,THz to $\sim 20$\,THz, limited on the low-frequency end by the need for large particles that fall out of suspension and at the high-frequency end by absorption bands from the polyimide aerogel. Prototype filters consistently show microwave indices of refraction of $n \leq 1.08$ with expected loss tangents of $\tan(\delta) \leq 0.001$, reducing signal loss due to in-band reflections and intrinsic material energy dissipation to negligible levels. 

\section{Backmatter}


\begin{backmatter}
\bmsection{Funding}
This work was supported by the National Aeronautics and Space Administration under grant numbers NNX14AB76A and NNH18ZDA001N-18-APRA18-0008, as well as the National Science Foundation Division of Astronomical Sciences for their support of this work under Grant Numbers 0959349, 1429236, 1636634, and 1654494. K.H acknowledges that the material is based upon work supported by NASA under award number 80GSFC21M0002. We also gratefully acknowledge support under the Goddard Internal Research and Development (IRAD) program.

\bmsection{Acknowledgments}
The authors gratefully acknowledge Aerogel Technologies, LLC for use of their facilities and support for sample fabrication during the initial stages of this work.

Portions of this work were presented at the SPIE conference on Astronomical Telescopes + Instrumentation in 2022, under papers titled ``Characterization of aerogel scattering filters for astronomical telescopes'' and ``Novel infrared-blocking aerogel scattering filters and their applications in astrophysical and planetary science observations.'' 

The authors declare no conflicts of interest.

Data underlying the results presented in this paper are not publicly available at this time but may be obtained from the authors upon reasonable request.

\bmsection{Supplemental document}
See Supplement 1 for supporting content.  

\newpage
\begin{center}
\begin{longtable}{|c|c|c|c|c|c|c|c|}
\caption*{Supplement 1. Chemical properties of filter samples fabricated as part of this work.}
\label{tab:Aerogel_Fab_run_table}\\
\hline
\multirow{2}{*}{Run} & Median particle & Diamond & \multirow{2}{*}{Density}& \multirow{2}{*}{Shrinkage} & \multirow{2}{*}{Porosity} & Surface & Thermal\\
{} & size & concentration & {} &{} & {} & area & conductivity \\
\# & [µm]& [mg/cm$^3$] & [mg/cm$^3$] & \% & \% & [m$^2$/g] & [mW/(m$\cdot$K)] \\ \hline
\endfirsthead
\multicolumn{8}{c}%
{{Supplement 1. -- continued from previous page}} \\ \hline
\multirow{2}{*}{Run} & Median particle & Diamond & \multirow{2}{*}{Density}& \multirow{2}{*}{Shrinkage} & \multirow{2}{*}{Porosity} & Surface & Thermal\\
{} & size & concentration & {} &{} & {} & area & conductivity \\
\# & [µm]& [mg/cm$^3$] & [mg/cm$^3$] & \% & \% & [m$^2$/g] & [mW/(m$\cdot$K)] \\ \hline
\endhead
\hline
\multicolumn{8}{|r|}{{Continued on next page}} \\ \hline
\endfoot
\hline
\endlastfoot

    1  & 0  & 0  & 107  & 6.48  & 92.5  & 480.98  & 28.9  \\
    \hline
    2  & 0  & 0  & 110  & 7.84  & 92.5  & -  & -  \\
    \hline
    3  & 4.2  & 20  & 136  & 8.95  & 91.1  & 392.6  & 24.8  \\
    \hline
    4  & 22.5  & 20  & 135  & 8.48  & 90.9  & 383.96  & 27.5  \\
    \hline
    5  & 47  & 20  & 136  & 8.77  & 91.3  & 298.93  & 30.9  \\
    \hline
    6  & 4.2  & 50  & 187  & 12.96  & 89.9  & 272.8  & -  \\
    \hline
    7  & 22.5  & 50  & 155  & 6.41  & 91.3  & 274.9  & 27.5  \\
    \hline
    8  & 47  & 50  & 161  & 8.77  & 91.7  & 325.3  & 27.7  \\
    \hline
    9  & 22.5  & 50  & 177  & 9.86  & 89.5  & 305.9  & 29.4  \\
    \hline
    10  & 9  & 50  & 177  & 13.07  & 89.9  & 179  & -  \\
    \hline
    11  & 22.5  & 50  & 168  & 7.86  & 89.9  & 228.6  & 24.8  \\
    \hline
    12  & 4.2  & 80  & 200  & 7.71  & 89.6  & 248.33  & 26.1  \\
    \hline
    13  & 22.5  & 80  & 201  & 8.18  & 89.4  & 179.33  & 27  \\
    \hline
    14  & 47  & 80  & 203  & 8.48  & 89.1  & 166.33  & 23.9  \\
    \hline
    15  & 0  & 0  & 108  & 7.84  & 91.7  & 479.6  & -  \\
    \hline
    16  & 0  & 0  & 109  & 7.7  & 92.5  & 467.06  & -  \\
    \hline
    17  & 0  & 0  & 109  & 8.62  & 92.3  & 448.45  & 27.4  \\
    \hline
    18  & 4.2  & 50  & 164  & 8.9  & 90.2  & 321.33  & -  \\
    \hline
    19  & 9  & 50  & 164  & 8.66  & 90.1  & 315.6  & 27.3  \\
    \hline
    20  & 15  & 50  & 163  & 9.26  & 89.8  & 317.9  & 28.5  \\
    \hline
    21  & 9  & 20  & 130  & 8.10  & 91.2  & 371.2  & 25.9  \\
    \hline
    22  & 15  & 20  & 130  & 8.08  & 91.4  & 375.6  & 24.9  \\
    \hline
    23  & 9  & 80  & 196  & 8.03  & 90.2  & 252.70  & 26.2  \\
    \hline
    24  & 9  & 20  & 132  & 8.7  & 91.3  & 385.4  & 27.2  \\
    \hline
    25  & 15  & 80  & 198  & 8.22  & 89.5  & 249.9  & 31.3  \\
    \hline
    26  & 0  & 0  & 117  & 10.61  & 91.5  & 472.4  & 31.5  \\
    \hline
    27  & 15  & 20  & 135  & 9.09  & 91.5  & 361.3  & 26.0  \\
    \hline
    28  & 30  & 80  & 206  & 9.59  & 85.2  & 230.4 & 34.8  \\
    \hline
    29  & 4.2  & 80  & 202  & 8.91  & 89.5  & 261.2 & 30.6  \\
    \hline
    \multirow{2}{*}{30} & 4.2 & 20 & \multirow{2}{*}{182} & \multirow{2}{*}{8.7} & \multirow{2}{*}{90.0} & \multirow{2}{*}{225.6} & \multirow{2}{*}{41.3} \\* 
    {} &47 &60 & {} & & & & \\*
    \hline 
    \multirow{2}{*}{31} & 4.2 & 20 & \multirow{2}{*}{223} & \multirow{2}{*}{14.1} & \multirow{2}{*}{89.2} & \multirow{2}{*}{180.0} & \multirow{2}{*}{26.9} \\*
    {} &47 &80 & {} & & & & \\*
    \hline
    \multirow{3}{*}{32} & 4.2 & 20& \multirow{3}{*}{201} & \multirow{3}{*}{9.1} & \multirow{3}{*}{90.1} & \multirow{3}{*}{196.7} & \multirow{3}{*}{25.1} \\*
    {} &15 &20 & {} & & & & \\*
    {} &47 &80 & {} & & & & \\*
    \hline
   \multirow{2}{*}{33} & 4.2 & 20 & \multirow{2}{*}{191} & \multirow{2}{*}{7.6} & \multirow{2}{*}{90.7} & \multirow{2}{*}{204.7} & \multirow{2}{*}{22.0} \\*
    {} &22.5 &60 & {} & & & & \\*
    \hline
    \multirow{2}{*}{34} & 4.2 & 40 & \multirow{2}{*}{196} & \multirow{2}{*}{8.3} & \multirow{2}{*}{90.4} & \multirow{2}{*}{200.4} & \multirow{2}{*}{26.5} \\*
    {} &15 &60 & {} & & & & \\* 
    \hline
\multirow{2}{*}{35}& 4.2 & 40 & \multirow{2}{*}{224} & \multirow{2}{*}{9.2} & \multirow{2}{*}{89.7} & \multirow{2}{*}{284.2} & \multirow{2}{*}{32.8} \\*
{} &15 &60 & {} & & & & \\* 
\hline
\multirow{2}{*}{36} & 9 & 20 & \multirow{2}{*}{193} & \multirow{2}{*}{10.2} & \multirow{2}{*}{90.3} & 309.2 & \multirow{2}{*}{29.6} \\*
{} &300 &60 & {} & & &177.8 & \\*
    \hline
\end{longtable}%
\end{center}

\end{backmatter}


\newpage
\bibliography{AerogelCharacterization_references}

\end{document}